\documentclass[12pt]{iopart}
\usepackage{iopams}  
\begin{document}\jl{1} 
\title{Quorum of observables for universal quantum estimation}
\author{G. Mauro D'Ariano, Lorenzo Maccone and Matteo G. A. Paris}
\address{{\sc Quantum Optics Group}, INFM at Dipartimento di Fisica 
`Alessandro Volta' \\ Universit\`a degli Studi di Pavia, via Bassi 6, 
I-27100 Pavia, Italy}
\begin{abstract} Any method for estimating the ensemble average of
arbitrary operator (observables or not, including the density matrix)
relates the quantity of interest to a complete set of observables, 
{\it i.e.}  a {\it quorum}. This corresponds to an expansion on an 
irreducible set of operators in the Liouville space.  We give two
general characterizations of these sets.  All the known unbiased
reconstruction techniques, {\it i.e.} ``quantum tomographies'', can be
described in this framework.  New operatorial resolutions are given 
that can be used to implement novel reconstruction schemes.
\end{abstract}
\section{Introduction}\label{s:in}
In order to characterize a quantum system, one can measure an
observable, or a set of observables, on repeated identical
preparations of the system. A question immediately arises: is this set
a {\it quorum}, {\it i.e.} is it sufficient to give a complete quantum
information on the system? In other words, is it possible to estimate
the expectation values of any system operator? This issue is of a
great practical interest for fundamental experiments in quantum
measurement theory, as well as for a potential application as a
quantum standard in the new field of quantum information
{\cite{popescu}}.\par Different estimation techniques have been
proposed tailored to different systems, such as the radiation field
{\cite{homtom}}, trapped ions and molecular vibrational states
{\cite{welschrev}}, spin systems {\cite{weigert}}, and a unified
approach is desirable. For set of operators that exhibit a group
symmetry, a general theory, the so called ``group tomography'', has
been established {\cite{grouptom}}.  However, not all estimation
techniques can be described within a group-theoretical scheme, and the
purpose of this paper is to give a more general framework. As we will
see, all the known quantum estimation techniques can be embodied in
the present approach. In addition, the formalism here presented is of
help for the derivation of new operatorial resolutions that result into 
novel estimation techniques.\par The paper is structured as
follows. In section {\ref{s:tom}}, we establish the general conditions
for a quorum of observables. Sections {\ref{s:qopt}}, {\ref{s:spin}},
and {\ref{s:fre}} are devoted to examples of the estimation technique,
for the harmonic oscillator, general spin systems, and the free
particle respectively. Section {\ref{s:outro}} closes the paper with a
summary.  We added {\ref{s:liouv}} to reformulate the theory in the
familiar Dirac formalism, and {\ref{s:gsm}} to give a constructive
algorithm to derive tomographic basis for finite-dimensional Hilbert
space.
\section{Quantum estimation}\label{s:tom}
The indirect (tomographic) reconstruction 
\footnote{\footnotesize
Among the existing estimation techniques for quantum systems, the
so-called quantum homodyne tomography of a single radiation mode has 
received much attention in the literature {\cite{welschrev,bilkent}}. 
In this paper, the term "tomography" is collectively used to denote
any kind of state-reconstruction technique.}  of an operator $A$ is
possible when there exists a resolution of the form
\begin{eqnarray}
\hat A = \int_{\cal X} dx \: \hbox{Tr} \left[ \hat A\: \hat B^\dagger 
(x)\right]\: \hat C(x)\;\label{gentomo},
\end{eqnarray}
where $x$ is a (possibly multidimensional) parameter living on a
(discrete or continuous) manifold ${\cal X}$. The only hypothesis
in (\ref{gentomo}) is the existence of the trace. The operators $\hat
C(x)$ are functions of the quorum of observables measured for the 
reconstruction, whereas the operators $\hat B(x)$ form the
{\it dual basis} of the set $\hat C(x)$. The term 
\begin{eqnarray}
{\cal E}[\hat A] (x) = \hbox{Tr} \left[ \hat A\: \hat B^\dagger (x)\right]\: \hat C(x)
\label{qest}\;
\end{eqnarray}
represents
the quantum estimator for the operator $\hat A$. Indeed, the expectation 
value of $\hat A$, namely the quantity of interest, is given by
the ensemble average
\begin{eqnarray} \fl
\langle \hat A\rangle\doteq\hbox{Tr}\left[\hat A\hat\varrho\right]=
 \int_{\cal X} dx \: \hbox{Tr} \left[ \hat A\: \hat B^\dagger (x)\right]\:
\hbox{Tr}\left[\hat C(x)\hat\varrho\right] \equiv \int_{\cal X} dx 
\langle{\cal E}[\hat A] (x) \rangle
\;\label{expval},
\end{eqnarray}
where $\hat\varrho$ is the density matrix of the quantum system under
investigation. The averaged estimator in Eq. (\ref{expval}) 
is the product of two terms: The quantity $\hbox{Tr}\left[\hat
C(x)\hat\varrho\right]$ depends only on the quantum state, and it is related 
to the probability distribution of the measurement outcomes, whereas the term 
$\hbox{Tr} \left[ \hat A\: \hat B^\dagger (x)\right]$ depends only on the 
quantity to be measured. In particular, the tomography of the quantum state 
of a system corresponds to writing Eq. (\ref{gentomo}) for the operators
$\hat A=|k\rangle\langle n|$, $\{|n\rangle\}$ being a given Hilbert space 
basis. For a given system, the existence of a set of operators $\hat C(x)$,
together with its dual basis $\hat B(x)$ allows universal quantum 
estimation, i. e. the reconstruction of any operator. \par
We now give two characterizations of the sets $\hat B(x)$ and $\hat C(x)$ 
that are necessary and sufficient conditions for writing
Eq. (\ref{gentomo}). \begin{quote} {\bf Condition 1: bi-orthogonality}
\\ \vspace{2pt}
\noindent Let us consider a complete orthonormal set $|n\rangle\:
(n=0,1,\cdots)$.  Formula (\ref{gentomo}) is equivalent to 
the bi-orthogonality condition
\begin{eqnarray}
\int_{\cal X} dx \: \langle q|\hat B^\dagger (x)|p \rangle \: 
\langle m|\hat C (x)|l \rangle \: = \: \delta_{mp} \delta_{lq}
\label{biort}\;,
\end{eqnarray}
where $\delta_{ij}$ is the Kronecker delta. Eq. (\ref{biort}) can be
straightforwardly generalized to a continuous basis. 
\end{quote}
\begin{quote}
{\bf Condition 2: completeness} \\ \vspace{2pt} \noindent
If the set of operators $\hat C (x)$ is {\em irreducible}, namely if 
any operator can be written as a linear combination of the $\hat C(x)$ as
\begin{eqnarray}
\hat A = \int_{\cal X} dx \: a(x) \: \hat C (x)
\;\label{irred},
\end{eqnarray}
then Eq. (\ref{gentomo}) is also equivalent to the trace condition
\begin{eqnarray}
\hbox{Tr}\left[\hat B^\dagger (x)\:\hat C (y)\right]\: = \: \delta (x,y)
\label{binorm}\;,
\end{eqnarray} 
where $\delta(x,y)$ is a reproducing kernel for the set $\hat B (x)$, namely
it is a function or a tempered distribution which satisfies
\begin{eqnarray}
\int_{\cal X}dx\;\hat B(x)\:\delta(x,y)=\hat B(y)
\;.\label{deltadef}
\end{eqnarray}
One can easily see that an analogous identity holds for the set of $C(x)$ 
\begin{eqnarray}
\int_{\cal X}dx\;\hat C(x)\:\delta(x,y)=\hat C(y)
\;\label{delta2}.
\end{eqnarray}
\end{quote}
The proofs are straightforward.  The irreducibility condition on the
operators $\hat C (x)$ is essential for the equivalence of
(\ref{gentomo}) and (\ref{binorm}). A simple counterexample is
provided by the set of projectors $\hat P (x) = |x\rangle\langle x|$
over the eigenstates of a selfadjoint operator $\hat X$. In fact,
Eq. (\ref{binorm}) is satisfied by $P(x)$. However, since they do
not form an irreducible set, it is not possible to express a generic
operator as $\hat O \neq \int_{\cal X} dx \: \langle x | \hat O | x
\rangle \; |x\rangle\langle x|$.
\par\noindent
If either the set $\hat B(x)$ or the set $\hat C(x)$ satisfy the
additional trace condition
\begin{eqnarray}
\hbox{Tr}\left[\hat B^\dag (y)\:\hat B (x)\right]\: = \: \delta (x,y)
\label{bbd1} \\
\hbox{Tr}\left[\hat C^\dag (y)\:\hat C (x)\right]\: = \: \delta (x,y)
\label{bbd}\;,
\end{eqnarray}
then we have $\hat C (x) = \hat B (x)$ (notice that neither $\hat
B(x)$ nor $\hat C(x)$ need to be unitary). In this case, Eq. (\ref{gentomo})
may be rewritten as
\begin{eqnarray}
\hat A = \int_{\cal X} dx \: \hbox{Tr} \left[ \hat A\: \hat C^\dagger
(x)\right]\:  
\hat C (x)
\label{gentomo2}\;.
\end{eqnarray}
\par
\par In abstract terms a certain number of observables $\hat Q_x$
constitute a quorum when there are functions $f_x(\hat Q_x)=\hat C(x)$
such that $C(x)$ form an irreducible set
Notice that if a set of observables $\hat Q_x$ constitutes a quorum, than
the set of projectors $|q\rangle_x {}_x\langle q|$ over their eigenvectors 
provides a quorum too, with the measure $dx$ in Eq. (\ref{gentomo}) including
the measure $dq$.
Of course it is of interest to connect a quorum of observables to a 
resolution of the form (\ref{gentomo}), since only in this case 
there can be a feasible reconstruction scheme. If a resolution formula 
is written in terms of a set of selfadjoint operators, the set itself 
constitutes the desired quorum. However, in 
general a quorum of observables is functionally connected to the 
corresponding resolution formula. If the operators $\hat C(x)$ are unitary, 
then they can always be considered as exponential of a set of selfadjoint 
operators, say $\hat Q_x$. The quantity $\hbox{Tr}\left[\hat C(x)
\hat\varrho\right]$ is thus connected with the moment generating function 
of the set $\hat Q_x$, and hence to the probability density $p(q;x)$ of the
measurement outcomes, which play the role of the Radon transform the 
quantum tomography  of the harmonic oscillator \cite{vgl}. 
Here, the basic resolution
formula involves the set of displacement operators $\hat D (\alpha) =
\exp [\alpha a^\dag - \bar\alpha a]$, which may be viewed as
exponential of the field-quadrature operators $\hat
x_{\phi}=\frac{1}{2} (ae^{-i\phi}+a^{\dag} e^{i\phi})$. 
In general, the operators $\hat C(x)$ can be any function (neither
self-adjoint nor unitary) of observables and, even more generally, 
they may be connected to POVMs rather than observables.  
In Sections \ref{s:kerr} and \ref{nonun} we will see examples 
of this situation.
\section{Quantum estimation for harmonic system}\label{s:qopt}
The harmonic oscillator (HO) model provides a detailed description of 
several systems of interest in quantum mechanics, as the vibrational 
states of molecules, the motion of an ion in a Paul 
trap, and a single mode radiation field. Different proposals have 
been suggested in order to reconstruct the quantum state of a 
harmonic system. As we will see in the following they can be summarized 
using the framework of the previous section, which is also useful 
for devising novel estimation techniques. 
\subsection{Quantum homodyne tomography}\label{ss:homtom}
Perhaps the most famous quantum estimation technique is given by the
so-called quantum homodyne tomography {\cite{tomo}}. Homodyne tomography 
applies to a single-mode radiation field as well as to
the vibrational state of a molecule or a trapped ion, and consists of
a set of repeated measurements of the quadrature operator $\hat
q_{\phi}=\frac{1}{2}(ae^{-i\phi}+a^{\dag} e^{i\phi})$ at different
values of the reference phase $\phi$
\footnote{\footnotesize For the vibrational tomography the quadrature
operator is a time-evolved position or momentum.}.  
For homodyne tomography, the relevant operatorial resolution is 
provided by the set of (irreducible) displacement operators 
$\hat D (\alpha) = \exp \left(\alpha a^\dag - \bar\alpha a\right), 
\alpha \in {\mathbb C}$. In fact, for the displacements Eqs. (\ref{binorm}) 
and (\ref{bbd}) hold since \begin{eqnarray}
\hbox{Tr}[\hat  D(\alpha)\hat D^\dag (\beta)] 
= \pi\delta^{(2)} (\alpha -\beta)\:,\;\label{biorthom}
\end{eqnarray}
and Eq. (\ref{gentomo2}) reduces to the Glauber formula 
\begin{eqnarray}
\hat A = \int_{\mathbb C} \frac{d^2\alpha}{\pi} \: \hbox{Tr} \left[ \hat A\: 
\hat D^\dag (\alpha )\right]\:  \hat D (\alpha ) \label{glauber}\;.
\end{eqnarray}
Changing to polar variables $\alpha = (i/2)k
e^{i\phi}$, Eq. \eref{glauber} becomes {\cite{bilkent}}
\begin{equation}
\hat A = \int^{\pi}_0\frac{d\phi}{\pi}\int^{+\infty}_{-\infty} 
\frac{d k\, |k|}{4}\,\hbox{Tr} (\hat A\: e^{ik\hat{q}_{\phi}})\, 
e^{-ik\hat{q}_{\phi}}\:, \label{op}
\end{equation}
which shows explicitly the dependence on the quorum $\hat q_\phi$.
After taking the ensemble average of both members, evaluating this
trace over the set of eigenvectors of $\hat{q}_{\phi}$, one obtains
\begin{eqnarray}
\langle \hat A  \rangle = 
\int^{\pi}_0\frac{d\phi }{\pi}  \int^{+\infty}_{-\infty}\!\! dq\: p(q;\phi) \: 
R[\hat A] (q;\phi) \label{qht1}\;,
\end{eqnarray}
where $p(q;\phi)={}_\phi\langle q|\hat\varrho|q\rangle_\phi$ is the
probability distribution of quadratures outcomes.  The tomographic
kernel for the operator $\hat A$ is given by \begin{eqnarray}
R[\hat A](q;\phi) =
\hbox{Tr} [\hat A\:K(q- \hat{q}_{\phi})]\;,\label{tomke}
\end{eqnarray}
where the integral kernel $K(z)$ is 
\begin{equation}
K(z) = -\frac{1}{2} \hbox{P} \frac{1}{z^2} \equiv
-\lim_{\varepsilon \rightarrow 0^+} \frac{1}{2} \hbox{Re}
\frac{1}{(z+i\varepsilon)^2}\,,
\label{kf}
\end{equation}
P denoting the Cauchy principal value. \par 
Using our condition
(\ref{binorm}) one can see that the Glauber formula can be generalized
to
\begin{eqnarray}
\hat A = \int_{\mathbb C} \frac{d^2\alpha}{\pi} \: \hbox{Tr} \left[ \hat A\: 
\hat F_1 \hat D (\alpha ) \hat F_2 \right]\:  \hat F_2^{-1}\hat D^\dag 
(\alpha )\hat F_1^{-1}
\label{genglauber}\;, 
\end{eqnarray}
where $\hat F_1$ and $\hat F_2$ are two generic invertible operators.
By choosing $\hat F^\dag_1 = \hat F_2 = \hat S (\zeta )$ with
\begin{eqnarray}
\hat S (\zeta) &=& \exp\left[\frac12 \left( \zeta^2 a^{\dag 2} - \bar{\zeta}^2
a^2\right)\right]\quad \zeta \in {\mathbb C}\label{rotsque}\;
\end{eqnarray}
the squeezing operator, we arrive to a different tomographic resolution
\begin{eqnarray}
\langle \hat A  \rangle = 
\int^{\pi}_0\frac{d\phi}{\pi}  \int^{+\infty}_{-\infty}\!\! dq 
\: p_{\zeta}(q;\phi) \: \hbox{Tr} \left[\hat A \: 
\hat K[q - \hat q_{\phi\zeta}]
\right] \label{qht5}\;,
\end{eqnarray}
in terms of the probability distribution of the generalized squeezed 
quadrature operators 
\begin{eqnarray}
\hat q_{\phi\zeta} &=& \hat S^\dag (\zeta)\:\hat
q_\phi\: \hat S (\zeta) \\  &=& 
\frac12 \left[ (\mu e^{i\phi}+\nu e^{-i\phi}) a^\dag+
(\mu e^{-i\phi}+\bar{\nu}e^{i\phi}) a \right]
\label{rotquad}\;,
\end{eqnarray}
where $\mu = \cosh |\zeta|$ and $\nu = \sinh |\zeta| \exp(2i
\arg[\zeta ])$.  Such an estimation technique has been investigated in
detail in Ref. {\cite{cometipare}}.
\subsection{Phase-space estimation techniques}
A different estimation technique based on the generalized Glauber formula
\eref{genglauber} may be obtained by putting $\hat F_1=\hat I$, the identity 
operator, and $\hat F_2=(-)^{a^\dag a}$, the parity operator. In this case 
one gets
\begin{eqnarray}
\hat A =  \int_{\mathbb C} \frac{d^2\alpha}{\pi} \: \hbox{Tr} \left[ 
\hat A \: \hat D^\dag ( \alpha) (-)^{a^\dag a}\right]\:(-)^{a^\dag a} 
\hat D (\alpha )\label{phspa0}\: ,\end{eqnarray} 
and therefore changing variable to $\alpha =2 \beta $ and using the relation 
$(-)^{a^\dag a} \hat D (2\beta ) = \hat D^\dag (\beta )(-)^{a^\dag a} \hat D (\beta )$ 
\begin{eqnarray}
\langle \hat A \rangle =  \int_{\mathbb C} \frac{d^2\beta}{\pi} 
\: \hbox{Tr} \left[\hat A\: 4 \hat D^\dag (\beta )(-)^{a^\dag a}\hat D (\beta)\right]\:
\hbox{Tr}\left[\hat D (\beta )\:\hat\varrho\:\hat D^\dag 
(\beta )\: (-)^{a^\dag a} \right]\label{phspa}\;.
\end{eqnarray}
Eq. (\ref{phspa}) says that it is possible to estimate a HO 
operator $\hat A$ by repeated measurement of the parity operator on displaced
versions of the state under investigation. An approximated implementation of
this technique for a single mode radiation field has been suggested
\cite{wodb,opa} through the measurement of the photon number probability on
signals displaced by means of a beam splitter. A similar schemes has
been used for the experimental determination of the motional quantum
state of a trapped atom \cite{leib}. The advantage of \eref{phspa}
compared to the approximated methods is in the possibility of directly
obtaining the kernel $K[\hat A](\alpha )$ for any operator
$\hat A$ for which the trace exists.  For instance, the
reconstruction of the density matrix in the Fock representation may
be obtained by averaging the kernel
\begin{eqnarray}
\fl 
K[|n\rangle\langle n+d||](\alpha ) &=& \langle n+d |4\: \hat D^\dag ( \alpha) (-)^{a^\dag a}
\hat D (\alpha )| n \rangle \\ &=& 4\: (-)^{n+d}\: \exp \{- 2 |\alpha |^2\} 
\sqrt{\frac{n!}{(n+d)!}} \: (2 \alpha  )^d \: L_n^d (4 |\alpha |^2)
\label{phspa1}\;, 
\end{eqnarray}
without the need of artificial cut-off in the Fock space \cite{leib}.  \par
\subsection{Nonlinear phase tomography}\label{s:kerr}
Let us now consider the set of selfadjoint operators given by
\begin{eqnarray}
\hat B(\psi ,\phi) &=& \hat V^\dag (\psi) \: \mu (\phi )\: \hat V (\psi)
\qquad\quad \psi\:, \phi \in [0,2\pi) \label{kerrT}\;, 
\end{eqnarray}
where $\mu (\phi)$ is the canonical London' POVM describing the ideal 
measurement of the HO phase and $\hat V (\psi)$ 
describes a nonlinear phase-shift, namely 
\begin{eqnarray}
\mu (\phi) &=& |e^{i\phi} \rangle\langle e^{i\phi} |
\nonumber \\ \hat V (\psi) &=& \exp 
\left[ i \: (a^\dag a)^2 \psi\right] \label{kerrop}\;,
\end{eqnarray}
where $|e^{i\phi}\rangle = \sum_n e^{in\phi} |n\rangle$ is the
Susskind-Glogower vector.
For a single-mode radiation field the action of $\hat V(\psi )$ corresponds 
to a nonlinear Kerr interaction, whereas for a trapped ion it could be 
obtained by laser excitation of vibronic levels \cite{kri}. \par 
One may argue that the measurement of the operators $\hat B(\psi ,\phi)$ 
provides a complete characterization of the state under investigation. 
This is indeed the case with $B(\psi,\phi)$ as a self-dual basis, 
as it can be easily proved using the bi-orthogonality condition \eref{biort}
\begin{eqnarray}\fl
\int_{0}^{2\pi}\int_{0}^{2\pi}\frac{d\phi}{2\pi}\frac{d\psi}{2\pi} \:
\langle m|\hat B(\psi ,\phi) |l \rangle \: \langle q|\hat B(\psi ,\phi) |p \rangle \:
= \delta_{p^2 - q^2,m^2 -l^2}\: \delta_{p-q,l-m} = \delta_{mp} \delta_{lq}
\label{kerrort}\;.
\end{eqnarray}
The kernel for the operator $\hat A$ is obtained from equation 
\eref{expval}
\begin{eqnarray}
K[\hat A ](\phi,\psi)= \hbox{Tr} \left[ \hat A \: \hat B(\phi,\psi ) \right]
\label{kerkerA}\;.
\end{eqnarray} 
In particular, the kernel $K[\hat P_{nd}](\phi,\psi)$ for the matrix
elements is given by  
\begin{eqnarray}
K[\hat P_{nd}](\phi,\psi)= \langle n+d | \hat B(\phi,\psi ) |n\rangle = 
\exp\left[ i \psi (d^2+2nd) + i \phi d\right]
\label{kerker}\;.
\end{eqnarray}
Notice that, for diagonal matrix elements, Eq. \eref{kerker} needs a
regularization procedure. In fact, the kernel for the projector
$\hat P_{nn}$ is given by 
\begin{eqnarray}
K_\varepsilon [\hat P_{nn}](\phi,\psi) =   
\exp\left[ i 2 \psi n \varepsilon   + i \phi \varepsilon \right]
\label{kerkerd}\;,
\end{eqnarray}
with the limit $\varepsilon \rightarrow 0$ that should be taken after the
average over the probability density $p(\phi,\psi) = \hbox{Tr}[\hat\varrho \:
\hat B (\phi,\psi )]$. A similar procedure should be employed for the
reconstruction of any operator which is a function of the number
operator $a^\dag a$ only.  \par The nonlinear "phase tomography"
presented in this section represents a novel resolution formula for
harmonic oscillator operators. The corresponding
reconstruction technique is based on an ideal phase measurement of the
Kerr-displaced state. Hence, in this scheme the quorum is a POVM.
\subsection{A nonunitary resolution formula}\label{nonun}
In this section we present a resolution formula in term of nonunitary operators
and show how its implementation would correspond to a generalized 
measurement i.e. a POVM.  The operators $\hat R_n(\phi)$ with  
$n \in {\mathbb Z}$ and $\phi \in [0,2\pi)$ are defined as follows
\begin{eqnarray}\fl
\hat R_n(\phi) = \left\{ \begin{array}{cr}
e_+^n \: e^{ia^\dag a \phi} & n \geq 0 \\
e_-^{-n} \: e^{ia^\dag a \phi} & n \leq 0
\end{array}\right.
\qquad
\hat R_n^\dag (\phi) = \left\{ \begin{array}{cr}
e^{-ia^\dag a \phi} \: e_-^n  & n \geq 0 \\
e^{-ia^\dag a \phi}\: e_+^{-n}  & n \leq 0
\end{array}\right.
\label{erre}\;,
\end{eqnarray}
where $e_- = \sum_n |n+1\rangle\langle n|$, $e_+ = \sum_n |n\rangle\langle n+1|$
are the so-called lowering and raising operators for the harmonic oscillator. 
The operators $\hat R_n(\phi)$ are not unitary, however $\hat R_n^\dag (\phi)$ plays 
the role of dual basis for $\hat R_n(\phi)$. This can be easily seen using
condition \eref{binorm}, for instance we have  
\begin{eqnarray}
\hbox{Tr}\left[\hat R_k^\dag (\psi)\: \hat R_n(\phi) \right] &\stackrel{n,k > 0}{=}&
\hbox{Tr}\left[e^{-ia^\dag a \psi} \: e_-^k \: e_+^n \: e^{ia^\dag a
\phi}\right] \nonumber \\ &=& \sum_p e^{ip(\phi-\psi)} \: \delta_{p+n,p+k} 
= \delta_{nk}\: \delta (\phi-\psi)\label{erretrace}\;,
\end{eqnarray}
and similarly for the other cases. 
As a consequence a generic operator $\hat A$ may be written as 
\begin{eqnarray}
\hat A = \sum_{n\in {\mathbb Z}} \int_{-\pi}^{\pi}\frac{d\phi}{2\pi}
\hbox{Tr}\left[ \hat A \:\hat R_n^\dag (\phi)\right] \hat R_n(\phi)
\label{erreres}\;.
\end{eqnarray}
For estimating HO operators one needs a
recipe to obtain the expectation value of $\hat R_n(\phi)$. Using a resolution of
identity in terms of phase vectors $|e^{i\phi}\rangle$
\begin{eqnarray}
\hat {\mathbb I} = \int_{-\pi}^{\pi}\frac{d\phi}{2\pi} \:
|e^{i \phi}\rangle\langle e^{i \phi} |
\label{idphase}\;
\end{eqnarray}
one evaluates the traces as 
\begin{eqnarray}\fl
\hbox{Tr}\left[\hat\varrho \: \hat R_q(\psi) \right]= \left\{
\begin{array}{cr}
\int_{-\pi}^{\pi}\frac{d\phi}{2\pi}\: e^{-iq\phi}\: \langle\phi -
\psi|\hat\varrho|\phi\rangle &q\geq 0\\         & \\ 
\int_{-\pi}^{\pi}\frac{d\phi}{2\pi}\: e^{iq(\phi+\psi)}\: \langle\phi 
|\hat\varrho|\phi+\psi\rangle &q\leq 0 \end{array}
\right. \label{erren}\;.
\end{eqnarray}
Eq. \ref{erren} implies that the knowledge of $\langle \hat
R_n(\phi)\rangle$ is equivalent to that of the density matrix in the
phase representation. 
\section{Quantum estimation in spin systems}\label{s:spin}
The recently born spin tomography {\cite{grouptom,weigert}}
allows to reconstruct the quantum state of a spin $s$ system.  It
employs measurements of the spin component in different directions,
{\it i.e.} it uses as quorum the set of operators $\vec S\cdot\vec n$,
where $\vec S$ is the spin operator and $\vec n$ a unit
vector. Various different quorums may be constructed by exploiting
different directions.\par
The easiest choice is to consider all possible directions. The
procedure to derive the tomographic formulas for this quorum is
analogous to the one employed in Sect. {\ref{ss:homtom}} for homodyne
tomography. The reconstruction formula for spin tomography for the
estimation of an arbitrary operator $\hat A$, using the measurement
outcomes $m$ of the component of the spin in all directions $\vec
n\doteq(\cos\varphi\sin\vartheta
,\sin\varphi\sin\vartheta,\cos\vartheta)$, is
\begin{eqnarray}
\langle\hat A\rangle=\sum_{m=-s}^s\int_{\Omega}\frac{d\vec
n}{4\pi}\;p(m,\vec n)\;R[\hat A](m,\vec n)\;\label{spintom},
\end{eqnarray}
where $p(m,\vec n)$ is the probability of obtaining the eigenvalue $m$
when measuring $\vec S\cdot\vec n$, $\hat R[\hat A](m,\vec n)$ is the
tomographic kernel for the operator $\hat A$, and $\Omega$ is the unit
sphere. In this case the operators $\hat C$ of Eq. ({\ref{gentomo})
are given by the set of projectors over the eigenstates $|m,\vec
n\rangle$ of the operators $\vec S\cdot\vec n$ for all directions
$\vec n$. Notice that it is a set of irreducible operators in the
system Hilbert space $\cal H$.  In order to find the dual basis $\hat
B$, one must consider the unitary operators obtained by exponentiating
the quorum, {\it i.e.} $\hat D(\psi,\vec n)=\exp(i\psi\vec S\cdot\vec
n)$, which satisfy the bi-orthogonality condition (\ref{biort}). In
fact, $\hat D(\psi,\vec n)$ constitutes a unitary irreducible
representation of the group SU(2), and the bi-orthogonality condition
is just the orthogonality relations between the matrix elements of the
group representation {\cite{murna}}, {\it i.e.} \begin{eqnarray}
\int_R dg\;{\hat D}_{jr}(g){\hat D}^\dagger_{tk}(g)=\frac
Vd\delta_{jk}\delta_{tr} 
\;\label{murnagh},
\end{eqnarray}
where $\hat D$ is a unitary irreducible representation of dimension
$d$, $dg$ is the group Haar invariant measure, and $V=\int_Rdg$. For
SU(2), with the $2s+1$ dimension unitary irreducible representation
$\hat D(\psi,\vec n)$, Haar's invariant measure is
$\sin^2\frac\psi2\sin\vartheta\: d\vartheta\: d\varphi\: d\psi$, and $\frac
Vd=\frac{4\pi^2}{2s+1}$. Thus, the bi-orthogonality condition is
\begin{eqnarray}
\frac{2s+1}{4\pi^2}\int_\Omega d\vec n
\int_0^{2\pi}d\psi\;\sin^2\frac\psi2\; \langle j|e^{i\psi\vec n\cdot\vec
S}|r\rangle\langle t|e^{-i\psi\vec n\cdot\vec
S}|k\rangle=\delta_{jk}\delta_{tr}
\;\label{murmaghsu2}, 
\end{eqnarray}
Using Eq. (\ref{murmaghsu2}) and condition 1, it is immediate to write
the spin tomography identity \begin{eqnarray}
\hat A=\frac{2s+1}{4\pi^2}\int_\Omega d\vec n 
\int_0^{2\pi}d\psi\;\sin^2\frac\psi2 \: \hbox{Tr} \left[\hat A\hat
D^\dagger(\psi,\vec n)\right]\hat D(\psi,\vec n)
\;\label{spinglaub}.
\end{eqnarray}
Notice the strict analogy between Eq. (\ref{spinglaub}) and Glauber's
formula (\ref{glauber}), analogy which derives from the group symmetry
that underlies both homodyne tomography (Weyl--Heisenberg group) and
spin tomography (SU(2) group). In fact, both these tomographies may be
derived in the domain of Group Tomography {\cite{grouptom}}. In order
to obtain the reconstruction formula (\ref{spintom}), one only has to
take the expectation value of both members of Eq. (\ref{spinglaub})
and to evaluate the expectation value trace on the eigenstates
$|m,\vec n\rangle$ of $\vec S\cdot\vec n$. Thus, the explicit form of
the tomographic kernel is obtained as
\begin{eqnarray}
R[\hat A](m,\vec
n)=\frac{2s+1}\pi\int_0^{2\pi}d\psi\;\sin^2\frac\psi2 \: \hbox{Tr}
\left[\hat A\; e^{-i\psi\left(\vec S\cdot\vec n-m\right)}\right]
\;\label{spinkern}.
\end{eqnarray}
\par 
As already anticipated, there are other possible quorums for spin
tomography. For example, for spin $s=\frac 12$ systems, a self-dual 
basis for the operator space is given by the Pauli matrices and the identity
$\{\frac 1{\sqrt{2}}\hat\sigma_x,\frac 1{\sqrt{2}}\hat\sigma_y,\frac
1{\sqrt{2}}\hat\sigma_z,\frac 1{\sqrt{2}}\hat 1\}$. Hence, we would
expect to find a tomographic identity where the above operators
constitute a quorum. In fact, from the property
$\hat\sigma_\alpha\cdot\hat\sigma_{\alpha'}=\hat
1\delta_{\alpha\alpha'}$ ($\alpha,\alpha'=x,y,z$), it is immediate to
see that both the bi-orthogonality relation (\ref{biort}) and the
trace condition (\ref{binorm}) hold. The following reconstruction
formula for an arbitrary $2\times 2$ matrix $\hat A$ derives
\begin{eqnarray}
\langle\hat A\rangle=\sum_{m=-\frac 12}^{\frac
12}\;\sum_{\alpha=x,y,z} p(m,\vec n_\alpha) \: \hbox{Tr}
\left[\hat A\hat\sigma_\alpha\right]m
+\frac 12 \: \hbox{Tr}
\left[\hat A\right]
\;\label{spintom12}.
\end{eqnarray}\par
In the case of generic $s$ spin system, Weigert has shown
{\cite{weigert}} that it is possible, by choosing $(2s+1)^2$ arbitrary
directions for $\vec n$, to obtain (in almost all cases) a quorum of
projectors $|s,\vec n_j\rangle\langle s,\vec n_j|$
($j=1,\cdots,(2s+1)^2$), where $|s,\vec n_j\rangle$ is the eigenstate
pertaining to the maximum eigenvalue $s$ of $\vec S\cdot\vec n_j$. The
search for the co-basis is done numerically, but since we have
a discrete quorum, it is possible to use the procedure derived from
Gram--Schmidt orthogonalization method, which is presented in
{\ref{s:gsm}}.
\section{Quantum state of a free particle}\label{s:fre}
Can we infer the state of a moving packet from position measurement in
time? The answer is positive as can be rigorously proved using our
condition \eref{biort}. The probability density of the position of a
free particle at the time $\tau$ is obtained from the selfadjoint
operator
\begin{eqnarray}
\hat R(x,\tau) = e^{- i\hat p^2 \tau}\: |x\rangle\langle x|\:
e^{i\hat p^2 \tau}
\label{free1}\;,
\end{eqnarray}
where $|x\rangle$ are eigenstates of the position operator and $\hat p$ the
momentum operator. We suppose for simplicity a particle with unit mass and 
use normalized unit $\hbar/2 =1$, such that the free Hamiltonian is given 
by $\hat H_F = \hat p^2$. The dual basis is constituted by the set of 
operator $\hat R(x,\tau)$ themselves, as follows from Eq. \eref{biort} and 
from the following relations ($| j\rangle$, $j=a,b,c,d$ denote position 
eigenstates)
\begin{eqnarray} \fl
\int_{\mathbb R}\int_{\mathbb R} &dx \:d\tau\: &
\langle a|\hat R(x,\tau)|b\rangle \: 
\langle c|\hat R(x,\tau)|d\rangle \nonumber \\ \fl &=& 
\int_{\mathbb R}\int_{\mathbb R} dx \:d\tau \:
e^{-i \tau (b^2-a^2 + c^2 -d^2)}\: 
\langle a|x \rangle \langle x|b\rangle\:
\langle c|x \rangle \langle x|d\rangle \nonumber \\ \fl &=&
\int_{\mathbb R}\int_{\mathbb R} dx\: d\tau \:
e^{-i \tau (b^2-a^2 + c^2 -d^2)}\: 
e^{i x (a -b + c -d )}\: \nonumber \\ \fl &=& \delta(a-c)\:\delta(b-d) 
\label{free2}\;.
\end{eqnarray}
Therefore, a generic free particle state can be written as 
\begin{eqnarray}
\hat \varrho = \int_{\mathbb R}\int_{\mathbb R} dx \:d\tau\: p(x,\tau) \: 
\hat R(x,\tau)\label{free3}\;
\end{eqnarray}
$p(x,\tau)= \hbox{Tr}[\hat\varrho\:\hat R(x,\tau)]$ being the probability
density of the particle to be at position $x$ at time $\tau$.
Eq. (\ref{free3}) can be generalized to particle moving in arbitrary 
potential, for details see Ref. \cite{thomas}.
\section{Summary}\label{s:outro}
A unified approach to quantum state estimation has been
presented. Some general conditions that guarantee that a set of
observables is sufficient for the estimation were given. A
constructive algorithm to derive dual basis, and therefore quantum
estimators, is suggested for the finite dimensional case. Our
framework allows to describe all known quantum tomographies ({\it
i.e.} unbiased state estimation procedures) for single mode radiation
field, trapped ions, vibrational states of molecules and spin
systems. Moreover, using our characterizations, some new tomographic
resolutions were derived. These may be used to implement novel
estimation techniques.
\appendix
\section{Liouville space formulation}\label{s:liouv}
In this section we reformulate the general scheme of quantum
estimation, given in Sect. {\ref{s:tom}}, by making use of the
properties of the linear space of operators acting on the system
Hilbert space $\cal H$, {\it i.e.}  the Liouville space ${\cal
L}({\cal H})$. This may results in a more convenient visualization of
of our conditions 1 and 2, since customary Dirac notation can be used. 
The Liouville space is built starting from the set of operators of 
Hilbert--Schmidt type, which is itself a Hilbert space, with the scalar 
product
\begin{eqnarray}
\langle\hat A|\hat B\rangle\doteq \hbox{Tr}\left[ \hat A^\dagger\:
\hat B\right] 
\;\label{scalprod}.
\end{eqnarray}
Using Dirac formalism on the operator Hilbert space, we find that kets 
are the operators of ${\cal L}({\cal H})$, {\it i.e.} $\hat O\doteq|\hat 
O\rangle$, while bras are obtained with the substitution $\hbox{Tr}[\;^
\centerdot\;\hat O^\dagger]\doteq\langle\hat O|\;^\centerdot\;$. 
The space of operators may be extended by considering non normalizable 
vectors in ${\cal L}({\cal H})$. Using Dirac notation, we
can rewrite all the formulas introduced so far. The generalized
tomography formula ({\ref{gentomo}}) corresponds to the expansion of
the vector $|\hat A\rangle$ on the non--orthogonal basis $|\hat
C(x)\rangle$, that will be referred to as ``spanning set'', and on its
dual $\langle\hat B(x)|$ ($x\in{\cal X}$), {\it i.e.}
\begin{eqnarray}
|\hat A\rangle=\int_{\cal X}dx\;
|\hat C(x)\rangle\langle\hat B(x)|\hat A\rangle
\;\label{tomonew}. 
\end{eqnarray}
In this framework, conditions 1 and 2 represent respectively the
identity resolution for the spanning set and the definition of the
dual vectors $\langle\hat B(x)|$ of the set $|\hat C(x)\rangle$. In
fact, we can readily rewrite Eqs. (\ref{biort}) and  (\ref{binorm}) 
as\footnote{\footnotesize Eq. (\ref{biortnew}) is readily obtained from
Eq. (\ref{biort}) by introducing the identity super-operator, defined
as $\hat{\hat 1}[\hat A]\doteq\hat A$ for any $\hat A\in{\cal L}({\cal
H})$. In fact, by using the basis $\{|n\rangle\}$ for $\cal H$, we see
that
\begin{eqnarray}
\langle k|\hat{\hat 1}\Bigl[|m\rangle\langle l|\Bigr]|p\rangle=
\delta_{mk}\delta_{lp}\nonumber 
\;\label{isuperop}.
\end{eqnarray}  }
\begin{eqnarray}
&&\int_{\cal X}dx\;|\hat C(x)\rangle\langle\hat B(x)|=\hat{\hat 1}
\;\label{biortnew},\\
&&\langle\hat B(x)|\hat C(y)\rangle=\delta(x,y)
\;\label{binormnew}.
\end{eqnarray}
\par 
The linear structure of the Liouville space provides a necessary and
sufficient condition to verify that a set of operators is a quorum. A
set of vectors $|\hat C(x)\rangle$, such that $\langle\hat B(x)|\hat
C(y)\rangle=\delta(x,y)\;\forall x$, is a spanning set (with
$\langle\hat B(x)|$ as dual set) iff the only operator $|\hat
O\rangle\in{\cal L}({\cal H})$ that is orthogonal to all $|\hat
C(x)\rangle$s is the null operator, {\it i.e.}  iff the two equivalent
conditions\begin{eqnarray}
\fl
\langle\hat O|\hat C(x)\rangle=\hbox{Tr}\left[\hat 
O^\dagger\hat C(x)\right]=0\qquad\qquad \langle\hat
B(x)|\hat O\rangle=\hbox{Tr}\left[\hat B^\dagger(x)\hat O\right]=0
\;\label{ortog1}
\end{eqnarray}
(for any $x\in{\cal X}$) imply that $\hat O=0$.  \par
In addition to the reconstruction of operators acting on the system
Hilbert space $\cal H$, one can extend the formalism also to the
reconstruction of super-operators acting on the system operator
space. A typical example is the Liouvillian super-operator that
evolves the system density operator into a density operator. In fact,
by introducing two resolutions of the identity in the operator space,
one can express any super-operator $\hat{\hat{\mbox{L}}}$ in terms of
its ``matrix elements'' $\langle\hat
B(x)|\hat{\hat{\mbox{L}}}|\hat C(y)\rangle$ on a basis for ${\cal L}({\cal
H})$, {\it i.e.}
\begin{eqnarray}
\hat{\hat{\mbox{L}}}=\int_{\cal X}dx\int_{\cal X}dy\;|\hat 
C(x)\rangle\langle\hat B(x)|\hat{\hat{\mbox{L}}}|\hat
C(y)\rangle\langle\hat 
B(y)|
\;\label{liouv}.
\end{eqnarray}
By taking a basis $\{|n\rangle\}$ of $\cal H$, Eq. (\ref{liouv})
rewrites as \begin{eqnarray}
L^{kl}_{mp}=\langle k|\hat{\hat{\mbox{L}}}\Bigl[|m\rangle\langle
l|\Bigl]|p\rangle=\nonumber\\
\int_{\cal X}dx\int_{\cal X}dy\;\hbox{Tr} \left[\hat 
C(x)|p\rangle\langle k|\right]\hbox{Tr} \left[\hat
B(y)|m\rangle\langle l|\right]\hbox{Tr} \left[\hat
B(x)\hat{\hat{\mbox{L}}}[\hat C(y)]\right]
\;\label{matrliouv}.
\end{eqnarray}
\section{Orthogonalization procedure}\label{s:gsm}
Here we give an algorithmic procedure, usable in the case of finite
quorums, to construct the set of dual operators $\langle \hat B_n|$ of
the quorum $|\hat C_n\rangle$. Using the Gram--Schmidt
orthogonalization method, one obtains a basis $|y_k\rangle$ from a
complete set of vectors $|C_k\rangle$ (assume for simplicity that all
$|C_k\rangle$ are non-zero and that in $\{|C_k\rangle\}$ there are no
couples of proportional vectors):
\begin{eqnarray}
\left\{\matrix{|y_0\rangle\doteq \frac1{N_0}\;|C_0\rangle\nonumber\\
|y_{k}\rangle\doteq
\frac1{N_k}\;\left(|C_{k}\rangle-\sum_{j=0}^{k-1}|y_j\rangle\langle 
y_j|C_{k}\rangle\right)} \right.
\;\label{gramschmidt},
\end{eqnarray}
where $N_0\doteq\parallel|C_0\rangle\parallel$ and
$N_k\doteq\parallel|C_{k}\rangle-\sum_{j=0}^{k-1}|y_j\rangle\langle
y_j|C_{k}\rangle\parallel$. Now, by writing the identity resolution
\begin{eqnarray}
\hat 1=\sum_{k=0}|y_k\rangle\langle y_k|\equiv\frac{|C_0\rangle}{N_0}\langle
y_0|+\sum_{k=1}\frac
1{N_k}\left(|C_k\rangle-\sum_{j=0}^{k-1}|y_j\rangle\langle
y_j|C_k\rangle\right)\langle y_k| \label{gsidres},
\end{eqnarray}
and using repeatedly Eq. (\ref{gramschmidt}) (expressing $|y_j\rangle$
of Eq. (\ref{gsidres}) in terms of the $|C_n\rangle$s and reorganizing
the terms) we can find the dual set $\langle B_n|$ as
\begin{eqnarray}
\langle B_0|=\frac {\langle y_0|}{N_0}-\frac{\langle
y_0|C_1\rangle\langle y_1|}{N_0N_1} +\left(-\frac{\langle
y_0|C_2\rangle}{N_0N_2}+\frac{\langle
y_0|C_1\rangle\langle y_1|C_2\rangle}{N_0N_1N_2}\right)\langle
y_2|+\cdots \nonumber\\
\langle B_1|=\frac{\langle y_1|}{N_1} -\frac{\langle
y_1|C_2\rangle\langle y_2|}{N_1N_2}+\left(-\frac{\langle
y_1|C_3\rangle}{N_1N_3}+\frac{\langle
y_1|C_2\rangle\langle y_2|C_3\rangle}{N_1N_2N_3}\right)\langle
y_3|+\cdots\nonumber\\ \cdots 
\;\label{gscob}
\end{eqnarray}
Unfortunately no such a general procedure appears to exist for the
case of a continuous spanning set. Many cases,though, satisfy the
conditions (\ref{bbd1}) and (\ref{bbd}), and hence we can write
$\langle\hat B(x)|=\left(|\hat C(x)\rangle\right)^\dagger$.
\ack This work has been partially supported by INFM
through project PAIS-1999-TWIN.

\end{document}